\documentclass[twocolumn]{article}
\usepackage[authoryear]{natbib}
\usepackage{geometry}
\usepackage{graphicx}
\usepackage{url}
\usepackage{array}
\usepackage{amsmath}
\usepackage{amsfonts}
\usepackage{amssymb}
\usepackage{mathrsfs}  
\newcolumntype{H}{>{\setbox0=\hbox\bgroup}c<{\egroup}@{}}

\begin{document}

\title{A possible tension between galaxy rotational velocity and observed physical properties}

\author{Lior Shamir and Darius McAdam \\ Kansas State University \\ Manhattan, KS 66506}


\date{}

\maketitle

\begin{abstract}

The discrepancy between the mass of galaxies and their rotational velocity is one of the most puzzling scientific phenomena. Despite over a century of research, this phenomenon is not fully understood. Common explanations include dark matter and MOND, among other theories. Here we report on another observation that shows tension between the physics of galaxy rotation and its rotational velocity. We compare the brightness of galaxies, and find that galaxies that spin in the same direction as the Milky Way have different brightness than galaxies that spin in the opposite direction. While such difference in brightness is expected due to Doppler shift, it is expected to be subtle. The results show that the difference in brightness is large enough to be detected by Earth-based telescopes. That observed difference corresponds to physical properties of galaxies with far greater rotational velocity than the rotational velocity of the Milky Way. The brightness difference is consistent in both the Northern galactic pole and the Southern galactic pole, and is not observed in parts of the sky that are perpendicular to the galactic pole. The differences are observed by several different instruments such DECam, SDSS, Pan-STARRS, and HST.  The observation is also consistent across annotation methods, including different computer-based methods, manual annotation, or crowdsourcing annotations through ``Galaxy Zoo'', all show similar results. 
Another possible explanation to the observation is parity violation in the large-scale structure, such that the magnitude of the parity violation was stronger in the earlier Universe. It can also be linked to other anomalies such as the $H_o$ tension. Analysis of $H_o$ using Ia supernovae shows smaller $H_o$ tension when the  spin directions of the host galaxies are consistent, although these results are based on a small number of supernovae, and may not be statistically significant.

\end{abstract}





\section{Introduction}
\label{introduction}

Despite over a century of research, the galaxy rotation curve anomaly  \citep{opik1922estimate,babcock1939rotation,oort1940some,rubin1970rotation,rubin1978extended,rubin1980rotational,rubin1985rotation,rubin2000} is still an unexplained observation. Early evidence that the galaxy rotation disagree with the physical properties of galaxies were observed as early as the first half of the 20th century \citep{slipher1914detection,wolf1914vierteljahresschr,pease1918rotation,babcock1939rotation,mayall1951structure}. In fact, the absence of Keplerian velocity decrease in the outer parts of galaxies was observed shortly after it became clear that galaxies are rotating objects \citep{rubin2000one}.

For instance, one of the most detailed early observations of the galaxy rotation curve anomaly was made by Jan Hendrik Oort, who analyzed the rotation and mass distribution of NGC 3115 \citep{oort1940some}. That work led to the conclusion that ``the distribution of mass in the system appears to bear almost no relation to that of light", and that ``the strongly condensed luminous system appears embedded in a large and more or less homogeneous mass of great density".

While that early work identified what is now considered the {\it dark matter halo}, preeminent astronomers of the time argued that the galaxy rotation was driven by Newtonian dynamics corresponding to the distribution of the visible light \citep{de1959general,schwarzschild1954mass}, which was based on the theory of that time. According to \cite{rubin2000one}, these opinions played a substantial role in ignoring the observations of the galaxy rotation curve anomaly, and led to adopting an incorrect Newtonian model as the physical model of galaxy rotation. Only several decades later the observations that galaxy rotation does not follow any known physical model was accepted by the ``mainstream'' astronomy community \citep{rubin2000one}.

After the tension between the galaxy rotation and its physical properties became an accepted observation, theoretical explanations were proposed. 
Perhaps the most common explanation to the anomaly is that the mass of the galaxy is dominated by dark matter \citep{zwicky1937masses} that does not interact with light or with any other known form of radiation \citep{rubin1983rotation}. While this explanation is widely accepted, there is still no direct conclusive evidence for the existence of dark matter \citep{sanders1990mass,mannheim2006alternatives,kroupa2012dark,kroupa2012failures,kroupa2015galaxies,arun2017dark,bertone2018new,skordis2019gravitational,sivaram2020mond,hofmeister2020debate}, and research efforts are still being continued. The dominance of the dark matter halo in spiral galaxies was also challenged by the properties of their rotation curves \citep{byrd2019ngc,byrd2021spiral}. The initial contention that the distribution of dark matter in the galaxy is constant \citep{oort1940some,donato2009constant} is in certain disagreement with its correlation with light and other galactic disk properties \citep{zhou2020absence}. Research efforts towards understanding the existence and nature of the contention that the mass of galaxies is dominated by dark matter are still being continued.

Another common paradigm related to the galaxy rotation curve anomaly is that galaxy rotation can be explained by modifications of the Newtonian Dynamics \citep{milgrom1983modification,milgrom2007mond,de1998testing,sanders1998virial,sanders2002modified,swaters2010testing,sanders2012ngc,iocco2015testing,diaz2018emergent,falcon2021large}. While dark matter became pivotal in standard cosmological models, the Modified Newtonian Dynamics (MOND) also showed agreement with observations \citep{kroupa2015galaxies,o2017alternative,wojnar2018simple,milgrom2019mond}, although other studies have shown tension between MOND predictions and data \citep{dodelson2011real,zhou2020absence}. Other explanations have also been proposed, such as \citep{sanders1990mass,capozziello2012dark,chadwick2013gravitational,farnes2018unifying,rivera2020alternative,nagao2020galactic,blake2021relativistic,larin2022towards}. But despite over a century of research, the physical nature of galaxy rotation is still a mystery, with no single proven and complete physical explanation.

\subsection{Previous reports on possible asymmetry in the distribution of galaxy spin directions}
\label{spin_direction_population}

While this study does not focus on the large-scale structure of the Universe, the observation reported here can be linked to certain observed anomalies in the large-scale structure reported in previous studies. One of these observations is the $H_o$ tension, as will be discussed in Section~\ref{H0}. Another observation is the proposed anomaly of asymmetry between the number spiral galaxies that spin in opposite directions, as observed from Earth. If the asymmetry reflects the large-scale structure of the Universe it might conflict with the standard cosmological theories. The results of this study propose that the observation is not necessary related to the large-scale structure, but to internal structure of galaxies. The relevance of the distribution of galaxy spin directions as observed from Earth to this study will be explained in Section~\ref{link}. The direct explanation of the observation described in this paper and the possible observed yet unexpected asymmetry between the number of galaxies spinning in opposite directions will also be discussed in detail in Section~\ref{link}.

The observation that the number of galaxies spinning towards one direction is not necessarily the same as the number of galaxies spinning in an opposite direction was noted as early as the mid 1980's, although with a relatively small number of galaxies \citep{macgillivray1985anisotropy}. The deployment of autonomous digital sky surveys allowed research with a much larger number of galaxies. Some experiments were based on manual annotations of the galaxies carried out by undergraduate students \citep{longo2011detection}, and other studies used automatic analysis of a large number of galaxies, as many as $\sim10^6$ galaxies \citep{shamir2022analysis2} or more \citep{shamir2022analysis} in a single experiment. The findings are consistent across different telescopes such as SDSS \citep{longo2011detection,shamir2012handedness,shamir2019large,shamir2020patterns,shamir2022possible}, Pan-STARRS \citep{shamir2020patterns}, HST \citep{shamir2020pasa}, DECam \citep{shamir2021large}, DES \citep{shamir2022asymmetry,shamir2022large}, and the DESI Legacy Survey \citep{shamir2022analysis}. Smaller scale studies showed correlation in spin directions \citep{lee2018study,lee2018wobbling}, also when the galaxies are too far from each other to have gravitational interactions \citep{lee2019galaxy,lee2019mysterious}, an observation that was defined as ``mysterious'' \citep{lee2019mysterious}.

\begin{figure*}
\centering
\includegraphics[scale=0.25]{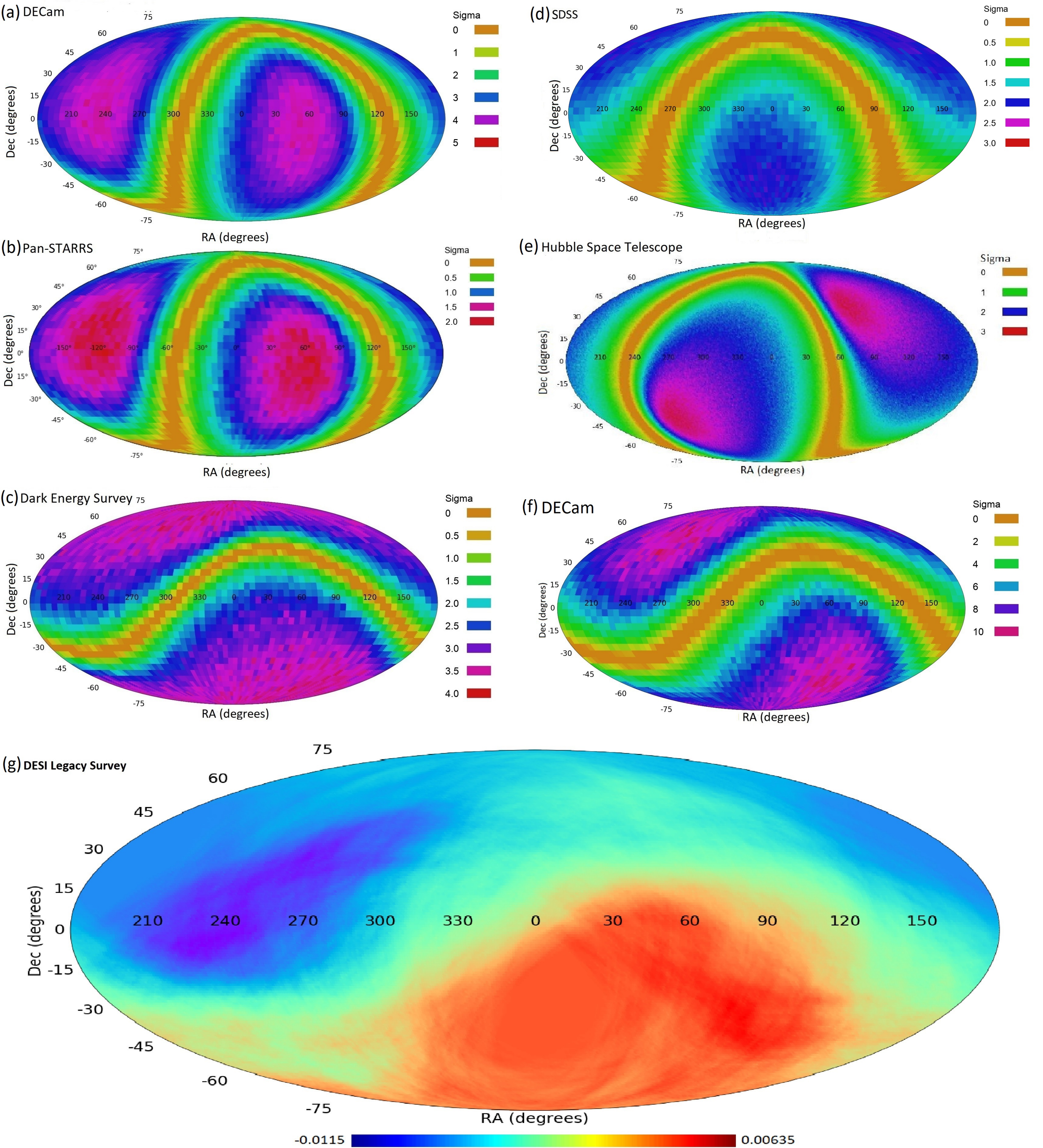}
\caption{Results of the analysis of dipole axes observed in different telescopes. The graphs show the probability to have a dipole axis by mere chance from each point in the sky. The analyses are taken from previous experiments \citep{shamir2020patterns,shamir2020pasa,shamir2021large,shamir2022analysis2,shamir2022asymmetry,shamir2022analysis}. The analyses shown in Panels (a) through (f) are based on statistical fitness of the distribution of spin directions into a dipole axis \citep{shamir2020patterns,shamir2020pasa,shamir2021large,shamir2022analysis2,shamir2022asymmetry}. The analysis shown in panel (g) is based on direct measurements of the difference in spin direction around each point in the sky.}
\label{all_telescopes}
\end{figure*}

According to that previous work, all telescopes and all analysis methods show a dipole axis exhibited by the spin directions of spiral galaxies \citep{shamir2022asymmetry,shamir2022large,shamir2022analysis}. That is, according to these reports the sky can be separated into two hemispheres, such that one hemisphere has a higher number of galaxies spinning clockwise, and the opposite hemisphere has a higher number of galaxies spinning counterclockwise. When using a large number of galaxies, a shifted window with $90^o$ radius can be used to deduce the asymmetry in all possible hemispheres \citep{shamir2022analysis}. Panel (g) in Figure~\ref{all_telescopes} shows the results of the asymmetry between the number of clockwise and counterclockwise galaxies in the hemisphere centered at each point in the sky, computed from a set of $\sim1.3$ million galaxies imaged by the DESI Legacy Survey. That experiment is described in detail in \citep{shamir2022analysis}, and its results show the asymmetry between hemispheres without an attempt to fit the spin directions into a certain statistical model. Fitting the spin directions into cosine dependence is shown in Panels (a) through (f), and suggest a statistically significant fitness into a dipole axis \citep{shamir2020patterns,shamir2020pasa,shamir2021large,shamir2022analysis2,shamir2022asymmetry}.

On the other hand, several studies suggested that the spin directions are random. These studies are discussed in \citep{shamir2022using} as well as \citep{shamir2022analysis,shamir2022analysis2,shamir2022asymmetry}. In summary, \cite{iye1991catalog} used $\sim6.5\cdot10^3$ galaxies and found no statistically significant non-random signal. As explained in \citep{shamir2022using}, the dataset was too small to show statistical significance, as showing a $2\sigma$ statistical significance in that manner requires at least $2.7\cdot10^4$ galaxies.

\cite{land2008galaxy} used SDSS galaxies annotated by anonymous volunteers, and showed a much higher number of galaxies spinning counterclockwise, which was the result of human bias in the annotation. After mirroring a small subset of these galaxies to correct for the human bias, the data showed 1.5\%-2\% more galaxies spinning counterclockwise. The small number of galaxies did not allow statistical significance (P$\simeq$0.13), but the direction and magnitude of that asymmetry is fully consistent with automatic analysis of SDSS galaxies in the same footprint \citep{shamir2020patterns}. 


An analysis of SDSS galaxies from the same footprint was done by applying automatic annotation using the {\it SpArcFiRe} algorithm \citep{Davis_2014}, and included a much higher number of galaxies \citep{hayes2017nature}. When applied to galaxies annotated as spirals by Galaxy Zoo 1, the asymmetry was statistically significant. Applying a machine-learning algorithm to annotate the galaxies showed random distribution, but that was observed after removing specifically all attributes that correlated with the asymmetry in galaxy spin directions. The removal of the attributes that correlate with the asymmetry naturally led to a symmetric dataset, as demonstrated experimentally in \citep{shamir2022using}.

\begin{figure}
\centering
\includegraphics[scale=0.24]{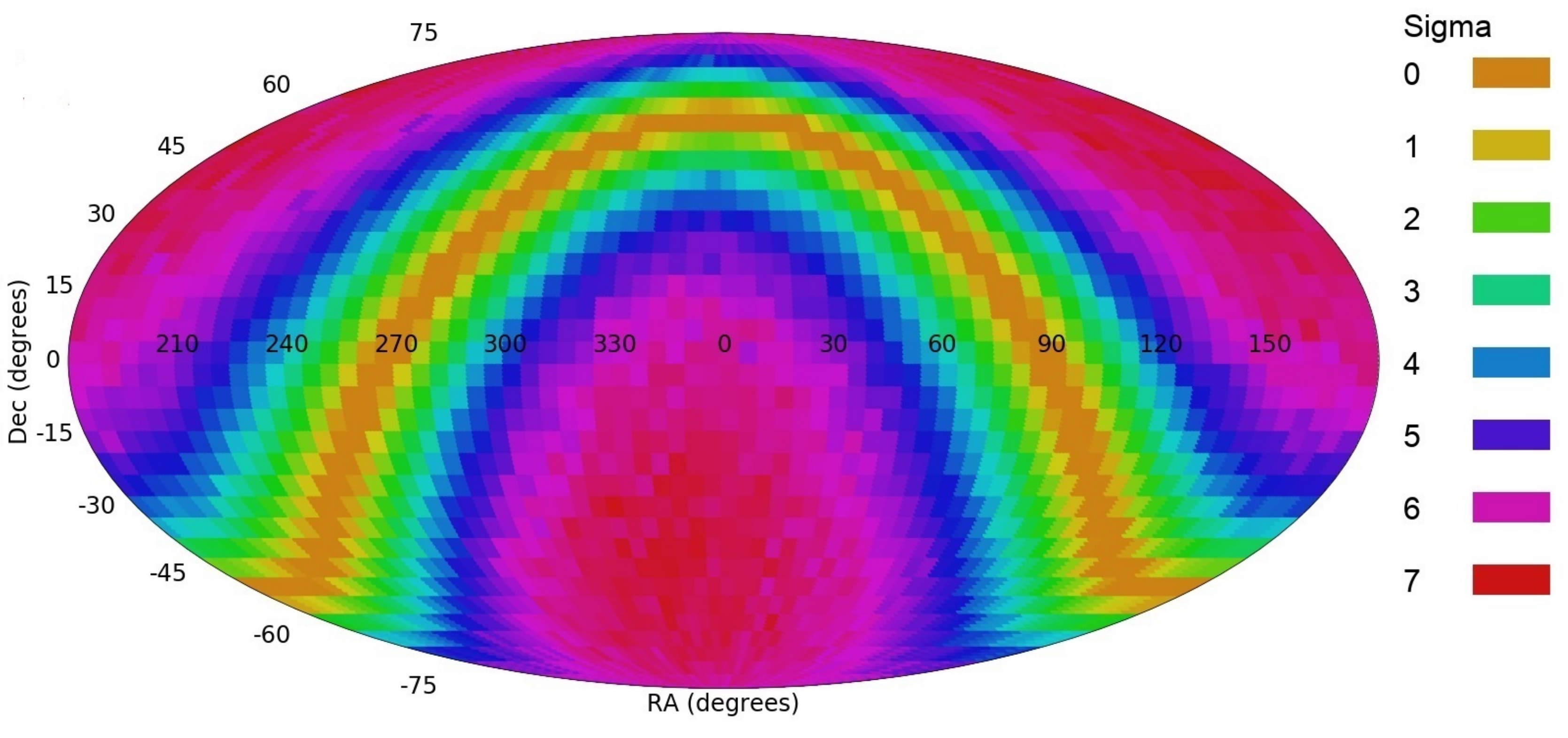}
\caption{Statistical strength of a dipole axis observed in galaxy spin directions in $\sim2.7\cdot10^5$ SDSS galaxies annotated by {\it SpArcFiRe}.}
\label{dipole_sparcfire}
\end{figure}

Repeating the experiment of \citep{hayes2017nature} by applying the same {\it SpArcFiRe} algorithm\footnote{https://github.com/waynebhayes/SpArcFiRe} to 666,416 SDSS galaxies used in Galaxy Zoo 1 without applying any form of pre-selection led to a dataset of 273,166 annotated galaxies. Figure~\ref{dipole_sparcfire} displays the statistical strength of a dipole axis in the galaxy spin directions in all parts of the sky. The Figure shows a dipole axis peaking at around $(\alpha=171^o, \delta=37^o)$, with statistical strength of 6.95$\sigma$. This experiment might not provide a complete proof to the existence of a dipole axis in galaxy spin directions, but it also does not conflict with previous experiments such as the results shown in Figure~\ref{all_telescopes}.

An interesting argument that the distribution of spin directions is random was made by \cite{iye2020spin}, who proposed that the observed asymmetry is the result of duplicate objects in the dataset used in \citep{shamir2017photometric}. However, no argument for the existence of any form of axis was made in \citep{shamir2017photometric}, and no argument for any kind of axis in that dataset was made in any other paper. More importantly, a simple analysis of that same ``clean'' dataset shows that it is not at all random \citep{shamir2022using}. For instance, Table~\ref{hemispheres} shows a very simple analysis of separating the exact same dataset used by \cite{iye2020spin} into two opposite hemispheres, showing that the distribution of spin directions form two hemispheres with inverse asymmetry \citep{shamir2022using}. Even if assuming that the distribution in the less populated hemisphere is random, a Bonferroni correction shows probability of $\sim0.01$. A Monte Carlo simulation shows probability of $\sim0.007$. Code and data to reproduce the analysis is available in \url{https://people.cs.ksu.edu/~lshamir/data/iye_et_al}.

\begin{table}
\scriptsize
\begin{tabular}{lcccc}
\hline
Hemisphere & \# cw          & \# ccw         &  $\frac{\#Z}{\#S}$  & P  \\ 
   (RA)         &                        &                         &                     & (one-tailed)  \\ 
\hline
$70^o-250^o$               & 23,037 & 22,442   &   1.0265   &  0.0026 \\ 
$>250^o \cup <70^o$   &  13,660 &  13,749  &   0.9935   &  0.29   \\ 
\hline
\end{tabular}
\caption{The distribution of spin directions in the ``clean'' dataset used in \citep{iye2020spin}.}
\label{hemispheres}
\end{table}

Analysis of a dipole axis in that dataset using Equation 2 in \citep{iye2020spin} shows a dipole axis with statistical significance of 2.14$\sigma$ \citep{shamir2022analysis}. That analysis is shown in Panel (d) of Figure~\ref{all_telescopes}. The exact same data used in \citep{iye2020spin}, the code that implements the analysis, step-by-step instructions for reproducing the analysis, and the output of the analysis are available in \url{https://people.cs.ksu.edu/~lshamir/data/iye_et_al}. More information is provided in \citep{shamir2022using,shamir2022analysis,shamir2022analysis2,shamir2022asymmetry}.

\subsection{Difference in brightness between galaxies spinning in opposite directions}
\label{brightness_difference}

A related observation is that galaxies that spin clockwise have, on average, different brightness than galaxies spinning counterclockwise. That is, in one hemisphere galaxies that spin clockwise are slightly brighter than counterclockwise galaxies, while in the opposite hemisphere galaxies that spin counterclockwise are brighter. 

Unlike the asymmetry between the number of galaxies spinning clockwise and the number of galaxies spinning counterclockwise, the difference in brightness of galaxies is a relatively new research question. The first report on the differences in colors between galaxies based on their spin direction was based on SDSS galaxies, with mild statistical significance \citep{shamir2013color}. That simple experiment included galaxies from SDSS with no separation to hemispheres or different parts of the sky. Further analysis showed differences in a large number of photometric measurements, but due to the large number of measurements tested no statistical significance was observed after applying a Bonferroni correction \citep{hoehn2014characteristics}.

Another experiment showed that by separating SDSS galaxies by their spin direction, a machine learning algorithm was able to predict the spin directions of the galaxies by their photometric attributes in accuracy far greater than mere chance \citep{shamir2016asymmetry}. That provided an indication that extra-galactic objects that spin in opposite directions are also photometrically different from each other. The dataset used in the experiment was too small to show statistically significant differences of specific measurements, but it provided early evidence of difference between the brightness of galaxies that spin clockwise, compared to the brightness of galaxies that spin counterclockwise. 

Separating into two hemispheres showed that in one hemispheres objects identified by the photometric pipeline as extended objects that spin clockwise are brighter, while in the opposite hemisphere objects that spin counterclockwise are brighter \citep{shamir2017large}. Similarly, color differences were also noted \citep{shamir2017colour}. The asymmetry was observed to peak very close to the galactic pole \citep{shamir2017large}. A similar observation was also made with a smaller number of HST galaxies \citep{shamir2020asymmetry}. The proximity of the peak of the asymmetry in galaxy brightness to the galactic pole led to the contention that the difference in brightness might be related to the rotational velocity of galaxies, and differences between galaxies that spin with the Milky Way compared to the brightness of galaxies that spin in the opposite direction \citep{shamir2017large,shamir2020asymmetry}. The difference was noticed not just from the data in the photometric pipelines of the sky surveys, but also in the analysis of the images \citep{sym14050934}. Explanations included anomalies in the large-scale structure, or in the internal structure of galaxies \citep{shamir2020asymmetry}.

\subsection{A possible link between brightness difference and the observed population of spiral galaxies with opposite spin directions}
\label{link}

Since every telescope has a limiting magnitude, the number of galaxies observed by that telescope is a direct function of the apparent brightness of the objects. Therefore, if in one hemisphere galaxies that spin clockwise are brighter than galaxies that spin counterclockwise, more galaxies spinning clockwise will be imaged, leading to a higher number of clockwise galaxies observed in that hemisphere. Similarly, brighter counterclockwise galaxies in the opposite hemisphere will lead to a higher number of galaxies spinning counterclockwise observed in that hemisphere. Therefore, an assumption that the brightness of a galaxy depends on its spin direction relative to the Milky Way can lead directly to an observed difference in the number of galaxies spinning in opposite directions.

Assuming that the brightness of a galaxy is related to its spin direction relative to the Milky Way, the difference between the number of clockwise and counterclockwise galaxies is expected to peak at around the galactic pole. The opposite side of the galactic pole is expected to show inverse asymmetry between galaxies with opposite spin directions, and that difference is expected to exhibit itself in the form of cosine dependence in other parts of the sky. As discussed in Section~\ref{spin_direction_population}, the asymmetry between the number of clockwise and counterclockwise galaxies can be fitted to cosine dependence with strong statistical significance, and form a dipole axis. Table~\ref{datasets} shows the most likely dipole axes determined in several different previous studies using several different telescopes.

\begin{table}
\scriptsize
\begin{tabular}{lccccc}
\hline
Paper       & Telescope	                          &  RA             & Dec              \\ 
reference   &                                               & (degrees)  & (degrees)         \\ 
\hline
 \citep{shamir2020patterns}     & SDSS                  &   229      &   -21          \\ 
 \citep{shamir2020patterns}     & Pan-STARRS           &   227      &      1        \\ 
 \citep{longo2011detection}      & SDSS                 &  217          & 32          \\ 
 \citep{shamir2016asymmetry}  & SDSS              &    165   &   30             \\ 
\citep{shamir2021particles}      & SDSS              &  165            &  40      \\ 
\citep{land2008galaxy}            & SDSS           &          161        &      11           \\			
\citep{shamir2012handedness}  & SDSS          & 132             &  32              \\ 
 \citep{shamir2022analysis}   & DESI & 243 & 39 \\
 \citep{shamir2021large} & DECam & 237 & 10 \\
  \citep{shamir2022asymmetry}  & DES & 255 & 47 \\
 \citep{shamir2022analysis2}      &    Combined   &  229    &  19  \\
\hline
\end{tabular}
\caption{The most likely dipole axis in the large-scale distribution of galaxy spin direction as reported in several different previous studies.}
\label{datasets}
\end{table}

\begin{figure*}
\centering
\includegraphics[scale=0.3]{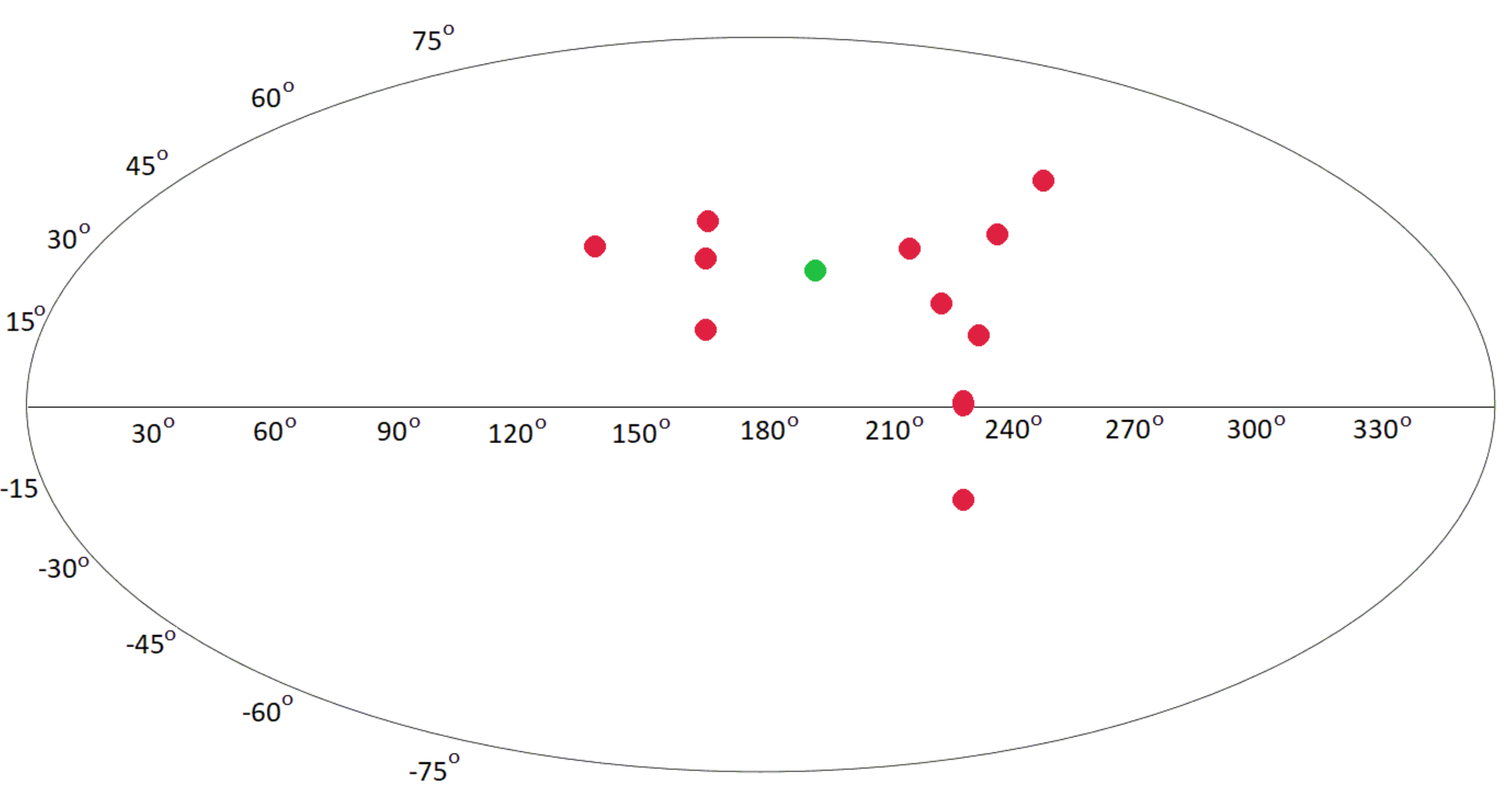}
\caption{The locations of the most likely dipole axes in several previous experiments specified in Table~\ref{datasets}, and the location of the galactic pole (green) at $(\alpha=192^o, \delta=27^o)$.}
\label{poles_position}
\end{figure*}

As the table shows, the location of the most likely dipole axes are in close proximity to the galactic pole at $(\alpha=192^o, \delta=27^o)$. That is shown visually in Figure~\ref{poles_position}. The locations of these axes are determined statistically, and therefore have certain statistical error. They are not aligned perfectly with the galactic pole, and the agreement can also be a coincidence. But the locations of these axes and their proximity to the galactic pole also do not conflict with the contention of a possible link between the galactic pole and the observation briefly mentioned in Section~\ref{spin_direction_population}, and will be discussed in much more details in the rest of this paper.

\subsection{Relativistic beaming and the link between galaxy brightness and its spin direction}
\label{relativistic_beaming}


Relativistic beaming has been studied in the context of high-velocity phenomena such as radio jets \citep{cohen2007relativistic,savolainen2010relativistic}, but was also analyzed statistically with quasars and radio galaxies \citep{orr1982relativistic,ishwara2000relativistic}, and a correlation has been reported between the flux and motion \citep{wills1986relativistic}. Interestingly, while it was found that relativistic beaming determines the observed flux in radio quasars \citep{padovani1992luminosity}, the observed statistical properties do not agree with the expected physical effect of relativistic beaming alone \citep{heinz2004constraints,odo2017relativistic}. It has been suggested that relativistic beaming can lead to radio source asymmetry \citep{alhassan2019relativistic}.

In addition to radio sources, relativistic beaming has been related to the light curve of active galactic nuclei \citep{lister2001relativistic}, black hole emission ring \citep{rummel2020constraining}, and binary black holes \citep{hayasaki2016detection}. It has also been proposed that relativistic beaming can lead to certain statistical asymmetry when observing a population of galaxies \citep{alam2017relativistic}.

According to relativistic beaming, a galaxy that rotates in the same direction as the Milky Way is expected to be have different brightness to an Earth-based observer compared to an identical galaxy that spins in the opposite direction. The brightness of a galaxy the brightness of its stars and other luminous objects, most of them are in a spin motion around the galaxy center.  A star or any other light-emitting object in a spiral galaxy rotating at velocity $V_r$ relative to a stationary observer will have a Doppler shift of its bolometric flux {\it F}. The expected observed flux {\it F} of a galaxy can be calculated by Equation~\ref{flux}

\begin{equation}
\label{flux}
F=F_o(1+4\cdot \frac{V_r}{c}),
\end{equation}
where $F_o$ is the flux of the luminous object if it was stationary relative to the observer, and $c$ is the speed of light \citep{loeb2003periodic,rybicki2008radiative}. Assuming that $\frac{v}{c}$ is $\sim$0.0007 as is approximately the case of the Sun in the Milky Way, a star rotating in the opposite direction compared to the Milky Way and observed on the galactic pole of the Milky Way will have $\frac{v}{c}$ of 0.0014 relative to an Earth-based observer. The $\frac{F}{F_0}$ of that star as observed from the Solar system is therefore $\simeq$1.0056. The brightness of a galaxy is driven by the total brightness of its luminous objects, and therefore the maximal expected difference between the magnitude of a face-on galaxy on the galactic pole that rotates in the same direction as the Milky Way and the magnitude of an identical galaxy spinning in the opposite way is $-2.5\log_{10}1.0056 \simeq 0.006$.

When the galaxy spins, its $F_o$ cannot be measured directly, and therefore $\frac{F}{F_o}$ cannot be determined observationally for a single galaxy. But when observing a large population of galaxies in the field around the galactic pole, the mean magnitude of the galaxies spinning in the same direction of the Milky Way can be compared to the mean magnitude of galaxies spinning in the opposite direction of the Milky Way. When a large number of galaxies is used, a statistically significant difference between the galaxy magnitude is expected. The purpose of this study is to compare the brightness of face-on galaxies spinning in the same direction as the Milky Way, and galaxies that spin in the opposite direction. That galaxies are imaged by the Dark Energy Camera (DECam) and other instruments at around the galactic pole.

The practice of using a large number of galaxies when a measurement is not possible with a single galaxy is a practice used in other tasks such as {\it weak lensing} \citep{van2000detection,wittman2000detection,mandelbaum2006galaxy,hirata2007intrinsic}. Here, the expected difference of 0.006 magnitude cannot be measured for a single galaxy, and therefore is measured as the mean magnitude of a large number of galaxies. To observe the maximum difference of 0.006 magnitudes, all galaxies need to be pure face-on galaxies, with no inclination, and all of them are expected to be exactly on the galactic pole of the Milky Way. These conditions are not practical, and therefore the expected observed difference is expected to be of less then 0.006 magnitudes.

\section{DECam Data}
\label{data}

The Dark Energy Camera (DECam) is placed on the Víctor M. Blanco 4m telescope in Cerro Tololo \citep{flaugher2012status}. Its footprint covers both the Northern and Southern galactic pole, allowing to compare them using the same instrument. The DECam data used in this study includes the two $60^o\times60^o$ fields centered around the Northern and Southern galactic pole. Additionally, two fields at $90^o$ from the galactic pole were used as control fields. The images were retrieved from the DESI Legacy Survey \cite{dey2019overview} server using the {\it Cutout} API. The images that were retrieved were images of objects identified as galaxies by the DESI Legacy Survey DR8 pipeline, and had magnitude lower than 19.5 in the g, r, or z band. Each image is a 256$\times$256 JPEG image, and the images were scaled using the Petrosian radius to ensure that the galaxy fits in the image. Because the objects are identified by the DESI Legacy Survey pipeline as extended objects, in some cases multiple objects can be part of the same galaxy. To ensure that each galaxy is represented once in the dataset, objects that have another object in the dataset within 0.01$^o$ are excluded from the dataset. The fields and the number of galaxies imaged by DECam in each field are specified in Table~\ref{fields}.

\begin{table}
\caption{The $(\alpha,\delta)$ coordinates of the centered and the number of galaxies in each $60^o\times60^o$ field.}
\label{fields}
\centering
\begin{tabular}{lccc}
\hline
Field      & \# galaxies  \\ 
center   &                     \\ 
\hline
$(192^o,27^o)$  &  1,309,498 \\ 
$(12^o,-27^o)$   &  6,376,803 \\ 
$(102^o,0^o)$   &   1,377,789 \\ 
$(282^o,0^o)$   & 1,266,036   \\ 
\hline
\end{tabular}
\end{table}

The galaxies were separated by their spin directions using the {\it Ganalyzer} algorithm \cite{shamir2011ganalyzer} as described in \cite{shamir2016asymmetry,shamir2017large,shamir2020patterns,shamir2017photometric,shamir2021large}. Ganalyzer is a deterministic model-driven algorithm that follows defined symmetric rules. It is not based on machine learning or complex data-driven rules that lead to non-intuitive classification schemes, and therefore its symmetric nature can be defined. The symmetricity of the algorithm ensures that the algorithm is not systematically biased, which is far more difficult to verify when using algorithms based on complex non-intuitive rules as typical in approaches such as deep neural networks.

In summary, the algorithm works by first converting each galaxy image into its radial intensity plot. The radial intensity plot of a galaxy image is a 35$\times$360 image, in which the pixel $(x,y)$ is the median value of the 5$\times$5 pixels around pixel coordinates $(O_x+\sin(\theta) \cdot r,O_y-\cos(\theta)\cdot r)$ in the original image, where {\it r} is the radial distance in percentage of the galaxy radius, $(O_x,O_y)$ is the center of the galaxy, and $\theta$ is the polar angle measured in degrees from the galaxy center. 

\begin{figure}
\centering
\includegraphics[scale=0.23]{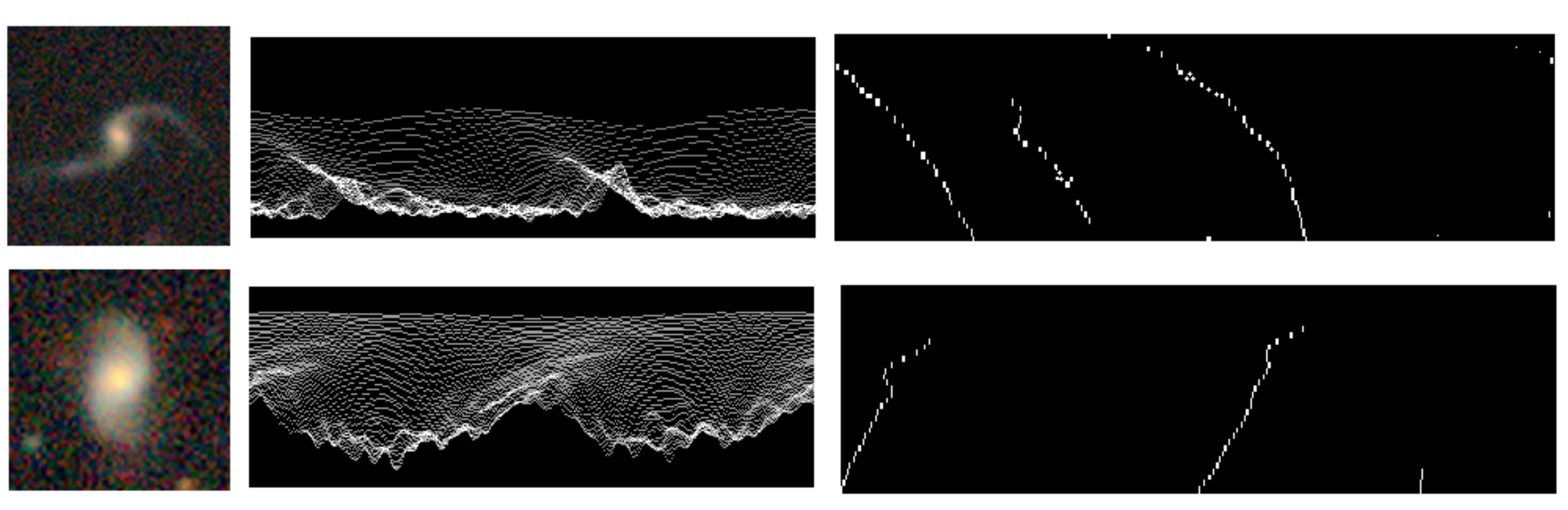}
\caption{Examples of galaxy images (left), the radial intensity plots of the galaxy images (center), and the peaks identified in the lines of the radial intensity plots (right). The direction of the lines generated by the peaks determines the direction of the arms.}
\label{radial_intensity_plot}
\end{figure}

Pixels on the arm of the galaxy are expected to be brighter than pixels at the same radial distance that are not on the arm, and therefore peaks in the radial intensity plot are expected to correspond to pixels on the arms of the galaxy. The arms can therefore be identified by applying a peak detection algorithm  \citep{morhavc2000identification} to the different lines in the radial intensity plot. After the peaks are identified, a linear regression is applied to the peaks in neighboring lines. The sign of the slope of the lines formed by the peaks reflects the direction of the curves of the arms, and consequently the spin direction of the galaxy \cite{shamir2016asymmetry,shamir2017large,shamir2020patterns,shamir2017photometric,shamir2021large}. Figure~\ref{radial_intensity_plot} shows an example of galaxy images, the radial intensity plots rendered from each galaxy image, and the peaks identified in the radial intensity plots.

Many galaxies are elliptical, and their spin directions cannot be determined. Other galaxies might not be elliptical, but still do not have identifiable direction of their spin. To reject galaxies that their spin direction cannot be determined, only galaxies that have three times more peeks in one direction compared to the other direction, and at least 30 identified peaks in their radial intensity plot \cite{shamir2016asymmetry,shamir2017large,shamir2017photometric,shamir2020patterns,shamir2021particles,shamir2021large}. The full description of the Ganalyzer algorithm is available in \cite{shamir2011ganalyzer,dojcsak2014quantitative,shamir2016asymmetry,shamir2017large,shamir2017photometric,shamir2020patterns,shamir2021large}. 

\begin{figure}[h]
\centering
\includegraphics[scale=0.4]{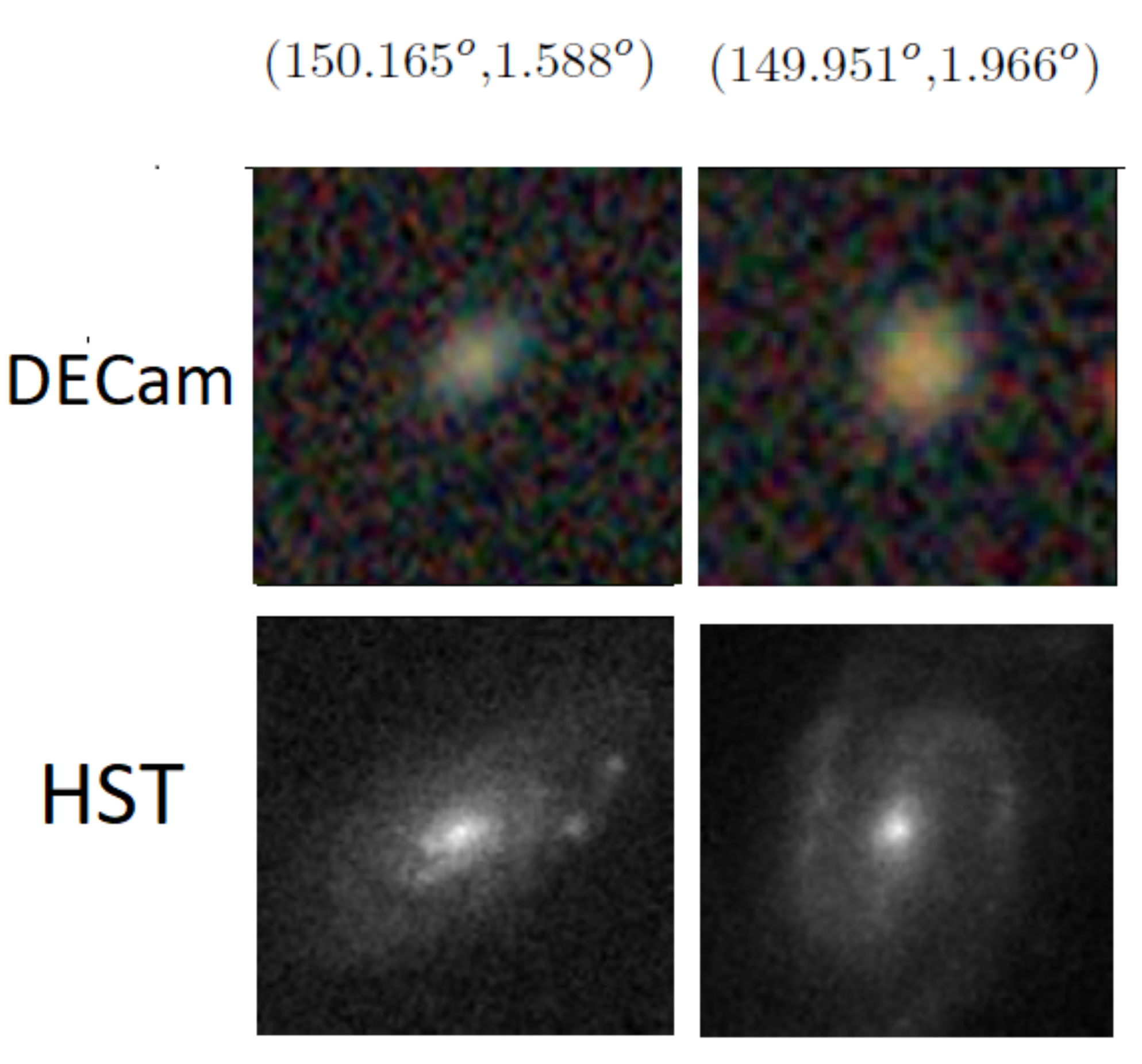}
\caption{Galaxies imaged by DECam and the same galaxies imaged by Hubble Space Telescope. The spin direction is clear in the HST images, but not clear in the DECam images.}
\label{sdss_vs_hst}
\end{figure}

The simple ``mechanical'' nature of {\it Ganalyzer} ensures that it is symmetric, as was also tested empirically in several previous studies. For instance, Figure 5 and Figure 9 in \citep{shamir2022asymmetry} show the results of the annotation after mirroring the galaxy images. A certain downside of using such analysis is that the annotation is not complete in the sense that all galaxies that have a spin direction are indeed annotated. Figure~\ref{sdss_vs_hst} shows an example of galaxies imaged by DECam, and the same galaxies imaged by Hubble Space Telescope (HST). As the figure shows, galaxies that have clear spin direction would be annotated as galaxies that do not have an identifiable spin direction, and therefore rejected from the analysis. Clearly, using HST images will also be subjected to incompleteness, as HST has its own limits on its imaging power. Since no telescope can provide a complete dataset in which the spin directions of all galaxies can be identified, the importance of the algorithm is its symmetric nature.

After separating the galaxy images by their spin directions, 400 random galaxies were observed, and none of these galaxies had a spin direction opposite to the spin direction it was annotated by the algorithm. While that does not ensure that the entire dataset does not have any wrongly annotated galaxies, it can be safely assumed that the dataset was into two subsets such that one subset has a much higher frequency of galaxies that spin clockwise, and the other subset has a much higher frequency of galaxies spinning counterclockwise. Another subset includes galaxies that their spin direction could not be determined, but due to the symmetric nature of the algorithm that subset is not expected to affect the ratio between the other two subsets. As will be shown in Section~\ref{results}, the inverse results in the opposite sides of the Galactic pole show no consistent bias of the algorithm.

\section{Results of DECam data}
\label{results}

Tables~\ref{north_pole} and~\ref{south_pole} show the difference between the mean magnitude of galaxies spinning in opposite ways in the fields around the North and the South galactic pole, respectively. To avoid potential erroneous magnitude values that can skew the mean magnitude, all magnitude values lower than 10 or greater than 25 were rejected from the analysis. 
The bands that were used in this study are the three optical bands of DESI Legacy Survey, which are the g, r, and z bands. Some galaxies do not have values for the flux in all bands, and that leads to a slightly different number of galaxies used in each band. 

The tables show the differences in the mean brightness between galaxies that spin in opposite directions at the field around the galactic poles. The tables show that the differences are statistically significant, as determined by the one-tailed P values of the Student t-test \citep{helmert1876genauigkeit}. While the differences in magnitude are observed in both hemispheres, the direction of the difference is inverse, such that clockwise galaxies are brighter in one hemisphere but dimmer i the opposite hemisphere.

The inverse difference in magnitude agrees with the expected difference caused by relativistic beaming in the two opposite sides of the galactic pole. It also shows that it is not caused by a bias of the annotation algorithm of photometric pipeline, as such bias is expected to be consistent across different fields, and is not expected to flip between the North and South galactic poles. 

\begin{table*}
\caption{The g, r, and z magnitude and the number of clockwise and counterclockwise galaxies in the $60^o \times60^o$ field centered around the North galactic pole $(\alpha=192^o,\delta=27^o)$.}   
\label{north_pole}
\centering
\begin{tabular}{lcccccc}
\hline
Band          & \# cw     & \# ccw    & Mag           & Mag  & $\Delta Mag$   & t-test P \\
                  &  galaxies & galaxies  &  ccw             &  cw  & \\                      
\hline
G &	20,918 & 21,253 &   20.06525$\pm$0.010  &   20.10073$\pm$0.010         & -0.03548  & 0.01	\\    
R   & 20,917  & 21,251   &  18.98522$\pm$0.008     &  19.01481$\pm$0.008       & -0.02958  &  0.01 \\  
Z   &  20,925 & 21,261   & 18.2934$\pm$0.007        &  18.31783$\pm$0.007        &  -0.02443  &  0.01 \\  
\hline
\end{tabular}
\end{table*}

\begin{table*}
\caption{The magnitude and number of clockwise and counterclockwise galaxies in the field centered around the South galactic pole $(\alpha=12^o,\delta=-27^o)$.}   
\label{south_pole}
\centering
\begin{tabular}{lcccccc}
\hline
Band          & \# cw     & \# ccw    & Mean           & Mean         & $\Delta Mag$   & t-test P \\
                  &  galaxies & galaxies  &Mag  ccw        & Mag  cw  & \\                     
\hline
G &	87,640 & 89,534  &   20.13622$\pm$0.004  &   20.11937$\pm$0.004    & 0.01685 	 &  0.003	\\   
R &  87,917  & 89,849  &    19.08793$\pm$0.003  & 19.07216$\pm$0.003 &  0.01574  & 0.0002  \\   
Z &  88,228 & 90,142    & 18.38424$\pm$0.003   & 18.37225$\pm$0.003         & 0.01199       &  0.0047 \\   
\hline
\end{tabular}
\end{table*}

Although the difference in magnitude around the galactic pole can be expected, in all cases the observed magnitude differences in both the Northern and Southern galactic poles are larger than the maximum expected difference in optimal conditions. As discussed in Section~\ref{relativistic_beaming}, the maximum expected magnitude difference expected under optimal conditions between galaxies at the galactic pole spinning in opposite ways is $\sim$0.006. For instance, the difference in the G band around the Northern galactic pole is $\sim$0.03548, which is larger than the expected maximum difference of 0.006. Because the standard error of the magnitude difference is $\sim$0.0142, the observed difference is larger by $\sim 2.1\sigma$ from the expected magnitude difference. From the six measurements in Tables~\ref{north_pole} and~\ref{south_pole}, the only measurement that does not meet the 0.05 P-value threshold for the difference between the expected and observed magnitude difference is the Z band of the Southern pole, with P$\simeq$0.08. Clearly, all measurements show strong statistical significance when assuming that the brightness of clockwise and counterclockwise galaxies around the galactic pole is the same. Because the galaxies in the Northern galactic pole are different galaxies than the galaxies in the South galactic pole, the two tests can be considered independent, and therefore the P value of the two tests is P$<0.0357\cdot0.08$, and therefore $P<0.0028$.

Tables~\ref{north_pole} and~\ref{south_pole} also show a difference between the number of galaxies that spin clockwise and the number of galaxies that spin counterclockwise. That difference agrees with the magnitude difference, as it is expected that if one type of galaxies is brighter to an Earth-based observer than the other type, more galaxies of that type will be identified. The previous reports on the large-scale differences between the number of galaxies spinning in opposite directions are specified in Section~\ref{spin_direction_population}, and their link to brightness differences is discussed in Section~\ref{link}. In summary, if one type of galaxies is indeed brighter to an Earth-based observer, the observed difference in the number of galaxies spinning in opposite directions can be the result of the difference in magnitude rather than the real population of spiral galaxies in the Universe.

The difference between the mean magnitude of galaxies spinning in opposite directions around the fields of the Northern and Southern galactic poles was compared to the two control fields that were selected as fields that are 90$^o$ away from the galactic pole. Tables~\ref{control_field1} and~\ref{control_field2} show the difference in magnitude around these fields. The tables use 47,017 and 41,244 galaxies, respectively. As both tables show, there is no statistically significant magnitude difference between the galaxies in these two fields. That shows that in 90$^o$ from the galactic pole, there is no observed difference between the brightness of galaxies spinning in opposite ways, as is the case for galaxies at around the galactic pole. That can be viewed as a link between the galactic pole and the differences in the brightness of galaxies spinning in opposite directions.

\begin{table}
\caption{The mean magnitude of clockwise and counterclockwise galaxies in the $60^o\times60^o$ window centered around the control field of $(\alpha=102^o,\delta=0^o)$.}   
\label{control_field1}
\centering
\scriptsize
\begin{tabular}{lHHcccccc}
\hline
Band          & \# cw     & \# ccw    & Mag           & Mag  & $\Delta Mag$   & P \\
                  &  galaxies & galaxies  &  ccw             &  cw  & \\                       
\hline
G &	23,709 & 23,308 &   20.16695$\pm$0.009  &   20.16628$\pm$0.009 & 0.000669	 &  0.96	\\
R   &  23,722 & 23,318  &  19.09924$\pm$0.007  & 19.10284$\pm$0.007  & -0.00356            &  0.713 \\
Z &  23,945 & 23,712    & 	18.39402$\pm$0.006 &  18.39436$\pm$0.006  & -0.00032          &   0.972 \\
\hline
\end{tabular}
\end{table}

\begin{table}
\caption{The mean magnitude of clockwise and counterclockwise galaxies in the $60^o\times60^o$ window centered around the control field of $(\alpha=282^o,\delta=0^o)$.}   
\label{control_field2}
\centering
\scriptsize
\begin{tabular}{lHHcccccc}
\hline
Band          & \# cw     & \# ccw    & Mag           & Mag  & $\Delta Mag$   & P \\
                  &  galaxies & galaxies  &  ccw             &  cw  & \\                 
\hline
G &	20,371 & 20873 &   20.21329$\pm$0.01  &   20.22103$\pm$0.01 & -0.00774	 &  $0.58$	\\
R   &  20,326 & 20,838  & 19.0787$\pm$0.008 &  19.0869$\pm$0.008  &  -0.0082 & 0.47 \\
Z &  20,883 & 21,387    & 18.37519$\pm$0.007	&  18.37565$\pm$0.007  &    -0.00045 & 0.96 \\
\hline
\end{tabular}
\end{table}

\section{Experiment with SDSS galaxies}
\label{sdss}

DECam provides images of a large number of galaxies, but until the Dark Energy Spectroscopic Instrument (DESI) sees first light most of these galaxies do not have spectra. To test a set of galaxies with spectra we used SDSS data. SDSS is inferior to DECam in its imaging capabilities, but as a mature redshift survey it collected spectra for a relatively high number of galaxies.

Images of 666,416 galaxies were downloaded by using the SDSS {\it cutout} API. The initial file format was 128$\times$128 JPEG, and each file was converted to the PNG format. These images were annotated by the {\it SpArcFiRe} (Scalable Automated Detection of Spiral Galaxy Arm) algorithm \citep{Davis_2014,hayes2017nature}. To test for consistency, the galaxy images were also mirrored by using the ``flip'' command of {\it ImageMagick}. That led to two annotated datasets, the first is the annotations of the original images, and the other is the annotations of the mirrored images. {\it SpArcFiRe} is an open source software with available source code\footnote{https://github.com/waynebhayes/SpArcFiRe}. {\it SpArcFiRe} is described in detail in \citep{Davis_2014}. The method works by identifying arm segments, and can then fit these segments to a logarithmic spiral arc. That allows {\it SpArcFiRe} to determine the spin direction of the galaxy. {\it SpArcFiRe} is a model-driven method, and is not based on machine learning that can lead to biases that are vgery difficult to detect \citep{dhar2022systematic}. 

The disadvantage of {\it SpArcFiRe} is that it has a certain error in the annotation, as also noted in Appendix A in \citep{hayes2017nature}. The {\it Ganalyzer} algorithm described in Section~\ref{data} allows to adjust the accuracy of the annotation by setting the minimum number of peaks required to make an annotation. If the number of peaks identified in the radial intensity plot is lower than the threshold, the galaxy is not used in the analysis. That leads to the rejection of a large number of galaxies. In the case of DECam, the initial number of galaxies is very high, and therefore even after rejecting a large number of galaxies the remaining galaxies make it a sufficiently large dataset to allow statistical analysis. {\it SpArcFiRe}, on the other hand, is far slower and has a certain error, but it also rejects a smaller number of galaxies. The use of {\it SpArcFiRe} also allows to use two different analysis methods.

\begin{figure}
\centering
\includegraphics[scale=0.85]{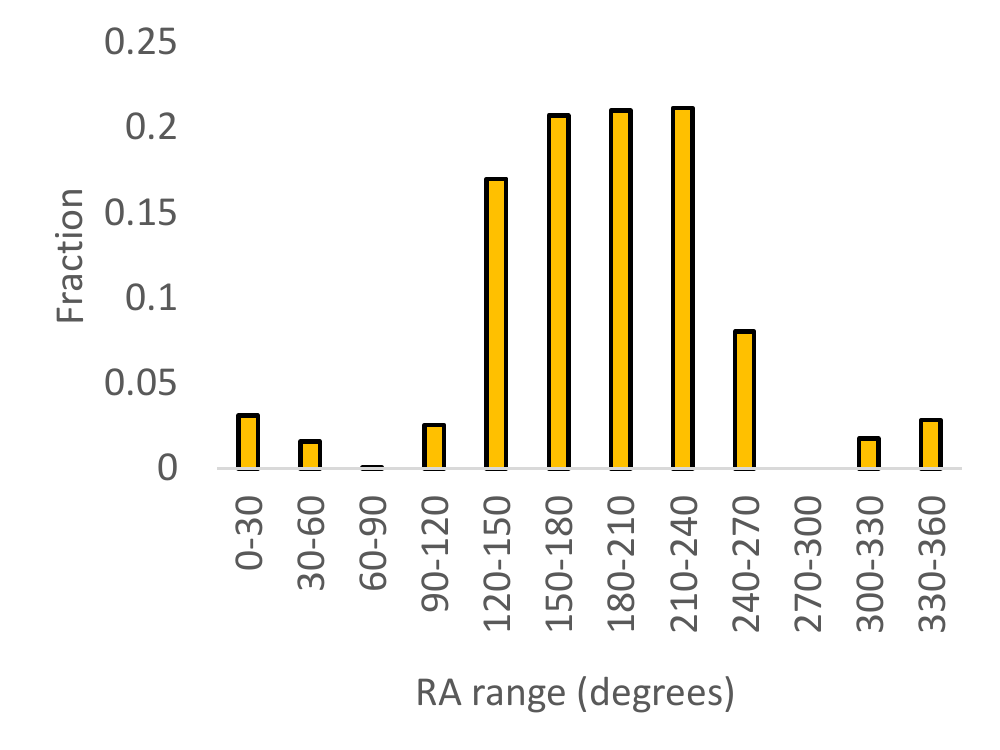}
\caption{The RA distribution of the SDSS galaxies.}
\label{ra_distribution_sdss}
\end{figure}

Classification of a single 128$\times$128 galaxy image requires $\sim$30 seconds when using a single core of a recent Intel Core-i7 processor. To reduce the response time, 100 cores were used to annotate the image data using {\it SpArcFiRe}. Figure~\ref{ra_distribution_sdss} displays the RA distribution of the galaxies. As the figure shows, the galaxy population is not distributed uniformly in the sky.

As the figure shows, the distribution of SDSS galaxies in the sky is not uniform. Fortunately for this specific study, the population of SDSS galaxies is relatively dense around the Northern galactic pole. That allows to study the difference in brightness of galaxies that spin with or against the spin direction of the Milky Way.

The annotation provided a set of 273,055 galaxies with annotated spin directions. The {\it SpArcFiRe} method was then applied again after mirroring the galaxy images, providing a set of 273,346 galaxies. The slight differences between the results after mirroring the images is mentioned in \citep{hayes2017nature}, and will be discussed later in this paper.

A first experiment was by just applying {\it SpArcFiRe} without any first step of selecting spiral galaxies. While the annotation of galaxies that are elliptical can add noise, it might be expected that the error in the annotation will be distributed equally between clockwise and counterclockwise galaxies. {\it SpArcFiRe} also does not force a certain spin direction, and can also annotate galaxies as not spinning in any identifiable direction. When {\it SpArcFiRe} is not able to identify the spin direction of the galaxy, . Table~\ref{sdss_all} shows the magnitude differences in the field of $60^o\times60^o$ around the Northern galactic pole. The annotation process provided 55,223 galaxies spinning counterclockwise, and 55,051 spinning clockwise in that field. Because {\it SpArcFiRe} is not fully symmetric, the experiment was repeated after mirroring all galaxy images, and the results are shown in Table~\ref{sdss_all_mirrored}. The annotation of the mirrored images provided a dataset of 55,874 mirrored galaxies spinning clockwise and 54,488 mirrored galaxies spinning counterclockwise. The list of galaxies and their annotations as assigned by {\it SpArcFiRe} is available at \url{https://people.cs.ksu.edu/~lshamir/data/sparcfire_data/}.

\begin{table}
\caption{The g, r, and z exponential magnitudes of SDSS galaxies that spin with the Milky Way and galaxies that spin in the opposite direction compared to the Milky Way in the field centered around the North galactic pole.}   
\label{sdss_all}
\scriptsize
\begin{tabular}{lHHcccccc}
\hline
Band          & \# cw     & \# ccw    & Mag           & Mag  & $\Delta Mag$   & P \\
                  &  galaxies & galaxies  &  cw             &  ccw  &                      &  t-test   \\                                    
\hline
G &	55,051 & 55,223 &   17.7652$\pm$0.004  &  17.7558$\pm$0.004         & 0.0094  & 0.035	\\
R   & 55,051  & 55,223  &  17.0016$\pm$0.003     &  16.9922$\pm$0.003       & 0.0094  & 0.013  \\
Z   & 55,051  & 55,223  & 16.449$\pm$0.003        & 16.4357$\pm$0.003        &  0.0132  &  0.001 \\
\hline
\end{tabular}
\end{table}

\begin{table}
\caption{The g, r, and z exponential magnitudes of mirrored SDSS galaxies that spin with the Milky Way and galaxies that spin in the opposite direction compared to the Milky Way in the field centered around the North galactic pole.}   
\label{sdss_all_mirrored}
\scriptsize
\begin{tabular}{lHHcccccc}
\hline
Band          & \# cw     & \# ccw    & Mag           & Mag  & $\Delta Mag$   & P \\
                  &  galaxies & galaxies  &  cw             &  ccw  &                      &  t-test   \\                                    
\hline
G &	55,874 & 54,488 &   17.7576$\pm$0.004  & 17.762$\pm$0.004         & -0.0044  & 	0.21 \\
R   & 55,874  & 54,488  &  16.9936$\pm$0.003     &  17.0016$\pm$0.003       & -0.008  & 0.025  \\
Z   & 55,874  & 54,488  & 16.4357$\pm$0.003        & 16.4479$\pm$0.003        &  -0.0122  &  0.002 \\
\hline
\end{tabular}
\end{table}

As the tables show, despite the certain inaccuracy of {\it SpArcFiRe}, the analysis still shows differences in the brightness of galaxies spinning clockwise and galaxies spinning counterclockwise at around the galactic pole. As expected, mirroring the galaxy images showed inverse results.

{\it SpArcFiRe} is designed to analyze spiral galaxies \citep{hayes2017nature}. To apply {\it SpArcFiRe} to spiral galaxies, a set of spiral galaxies was separated from the other galaxies by using the {\it Ganalyzer} method \citep{shamir2011ganalyzer}. As a model-driven method, the analysis is not based on any kind of machine learning, and therefore it is not subjected to possible biases in the training data or the learning process \citep{dhar2022systematic}. The simple ``mechanical" nature of {\it Ganalyzer} allows it to be fully symmetric \citep{shamir2021large,shamir2022asymmetry}. 

Table~\ref{north_pole_sdss} shows the number of clockwise and counterclockwise galaxies in the SDSS data after selecting the spiral galaxies, limited to the $60^o \times 60^o$ part of the sky centered at the Northern galactic pole. That dataset contained 27,196 galaxies spinning clockwise and 27,671 galaxies spinning counterclockwise. Since SDSS footprint covers mostly the Northern hemisphere, the Southern galactic pole is outside of its footprint. According to Table~\ref{north_pole_sdss}, the results shows statistically significant differences between the brightness of galaxies that spin in opposite directions.

\begin{table}
\caption{The g, r, and z exponential magnitudes of SDSS galaxies annotated as spiral galaxies. The analysis is limited to galaxies in the $60^o\times60^o$ centered at the Northern galactic pole.}   
\label{north_pole_sdss}
\scriptsize
\begin{tabular}{lHHcccccc}
\hline
Band          & \# cw     & \# ccw    & Mag           & Mag  & $\Delta Mag$   & P \\
                  &  galaxies & galaxies  &  cw             &  ccw  &                      &  t-test   \\                                    
\hline
G &	27,196 & 27,671 &   17.7095$\pm$0.005  &  17.6948$\pm$0.005         & 0.0147  & 0.0376	\\
R   & 27,197  &  27,671  &  16.9893$\pm$0.004     &  16.9745$\pm$0.004       & 0.0148  &  0.0089 \\
Z   & 27,196  &  27,671  & 16.4564$\pm$0.004        & 16.4393$\pm$0.004        &  0.0171  &  0.0025 \\
\hline
\end{tabular}
\end{table}


The SDSS galaxies used in this study have redshift, which allows to separate them to redshift ranges. Table~\ref{north_pole_sdss_z007} shows the same analysis, but when limiting the galaxies to $z<0.07$. When limiting the redshift to 0.07, the dataset included 8,409 clockwise galaxies and 8,748 counterclockwise galaxies. The results show a larger difference in magnitude in that redshift range. That, however, is not necessarily of an astronomical meaning, and could be linked to the better ability of {\it SpArcFiRe} to annotate galaxies at lower redshift ranges. As the other experiments with SDSS galaxies, the results are in agreement with the results of the DECam galaxies shown in Section~\ref{results}.



\begin{table}
\caption{The g, r, and z magnitudes of SDSS galaxies limited to $z<0.07$ that spin in the opposite directions in the field centered around the North galactic pole.}   
\label{north_pole_sdss_z007}
\scriptsize
\begin{tabular}{lHHcccccc}
\hline
Band          & \# cw     & \# ccw    & Mag           & Mag  & $\Delta Mag$   & P \\
                  &  galaxies & galaxies  &  cw             &  ccw  &                                        &  t-test \\ 
\hline
G &	8,409  & 8,748 &   16.9437$\pm$0.009      &  16.9192$\pm$0.009         & 0.0245  & 0.027 \\
R   & 8,409  &  8,748  &  16.4017$\pm$0.009     &  16.3723$\pm$0.009       &  0.0294 &  0.01 \\
Z   & 8,409  &  8,748  & 15.9827$\pm$0.009      & 15.9499$\pm$0.009        & 0.0328   &  0.005 \\
\hline
\end{tabular}
\end{table}


\subsection{Analysis with crowdsourcing data from {\it Galaxy Zoo 1}}
\label{galaxy_zoo}

One of the previous attempts to annotate galaxies by their spin direction was done by crowdsourcing through the {\it Galaxy Zoo 1} project \citep{lintott2008galaxy}. According to Galaxy Zoo 1, anonymous volunteers used a web-based user interface to manually annotate galaxies by their shape. Among other features, the users were asked to annotate the spin direction of the galaxies. While not all annotations are expected to be correct, the majority of the votes is expected to provide a certain indication regarding the spin direction.

The Galaxy Zoo annotations can be used to perform an experiment similar to the experiments described above, but with galaxies annotated manually by a large number of volunteers. Fortunately, the $60^o \times 60^o$ part of the sky centered around the Northern galactic pole is fairly populated by galaxies that were annotated through Galaxy Zoo 1, and the majority of Galaxy Zoo annotated galaxies are concentrated around that part of the sky. That allows to obtain the profile of brightness differences between galaxies that were annotated by Galaxy Zoo to spin clockwise, compared to the brightness of Galaxy Zoo galaxies that were annotated as spinning counterclockwise.

The Galaxy Zoo 1 annotations were taken from the ``zooVotes'' table of SDSS DR8. The galaxies include only galaxies of which 60\% or more of the voters agreed on their spiral nature and spin direction. That is, the field ``p\_cw'' was greater or equal to 0.6 for galaxies that spin clockwise, and the field ``p\_acw'' was greater or equal to 0.6 for galaxies that spin counterclockwise. As before, missing values or flag values were removed. That provided a dataset of 11,150 galaxies spinning clockwise, and 11,907 galaxies spinning counterclockwise. All galaxies are inside the $60^o \times 60^o$ region centered at the Northern galactic pole. Table~\ref{galaxy_zoo06} shows the brightness differences in the g, r, and z bands for the Galaxy Zoo galaxies. As the table shows, the differences between the brightness of galaxies spinning in opposite directions in the Northern galactic are noticeable also in the galaxies annotated by Galaxy Zoo. The corresponding part of the sky in the Southern galactic pole has merely a total of 2,005 annotated galaxies, and therefore analysis of the Southern galactic pole is not possible in the same manner it was done with the DECam data.
  
\begin{table}
\caption{The g, r, and z exponential magnitudes of SDSS galaxies annotated by Galaxy Zoo in the field centered around the North galactic pole.}   
\label{galaxy_zoo06}
\scriptsize
\centering
\begin{tabular}{lcccccc}
\hline
Band        & Mag           & Mag  & $\Delta Mag$   & P \\
               &  cw            &  ccw  &                      & t-test  \\
\hline
G  &  16.9765$\pm$0.01      &  16.9579$\pm$0.01         & 0.0186  & 0.09 \\
R  &  16.4129$\pm$0.01     &  16.3723$\pm$0.01       & 0.0406  &  0.002 \\   
Z  &  15.9817$\pm$0.01      & 15.9539$\pm$0.01        & 0.0278   & 0.025 \\  
\hline
\end{tabular}
\end{table}

The {\it Galaxy Zoo 1} defines a ``superclean'' annotation as an annotation that 95\% of the annotators provided the same annotation. Table~\ref{galaxy_zoo95} shows the differences in brightness of galaxies in the $60^o \times 60^o$ region around the Northern galactic pole such that all annotations of the galaxies meet the ``superclean'' criterion. The requirement for higher agreement of the annotators can lead to cleaner annotations, but that comes on the expense of the size of the dataset. When using just galaxies on which 95\% of the annotators agree, the total number of annotated galaxies is merely 4,065. The number of clockwise galaxies is 1875, and the number of galaxies annotated as spinning counterclockwise is 2190. The results also show that galaxies spinning counterclockwise around the North galactic pole are brighter. The results are not statistically significant, which can be attributed to the smaller number if galaxies that satisfy the higher annotation agreement threshold. But the difference observed in the ``superclean'' galaxies also does not conflict with the difference observed with the larger datasets.

\begin{table}
\caption{The g, r, and z magnitudes of SDSS galaxies annotated as ``superclean'' by Galaxy Zoo in the field centered around the North galactic pole.}   
\label{galaxy_zoo95}
\scriptsize
\centering
\begin{tabular}{lcccccc}
\hline
Band        & Mag           & Mag  &  $\Delta Mag$   & P \\
               &  cw            &  ccw  &                        &  t-test \\  
\hline
G  &  16.3496$\pm$0.03      &  16.3122$\pm$0.03         & 0.0374  & 0.189 \\
R  &  15.7912$\pm$0.03     &  15.7443$\pm$0.03       & 0.0468  &  0.135 \\
Z  &  15.377$\pm$0.03      & 15.3235$\pm$0.03        & 0.0535   & 0.104 \\
\hline
\end{tabular}
\end{table}

The Galaxy Zoo annotations are known to be subjected to certain biases that are often difficult to fully profile \citep{hayes2017nature}. The brightness differences observed with galaxies annotated by Galaxy Zoo can therefore be the results of some unknown bias where the volunteers tend to annotate brighter galaxies as galaxies rotating counterclockwise. These biases can be difficult to quantify and fully profile, and therefore the results when using the Galaxy Zoo annotations might  not be sufficient to provide strong evidence of brightness differences. The agreement between the results of Galaxy Zoo and the results of DECam and SDSS can also be considered a coincidence. But the results of Galaxy Zoo also do not conflict with the results shown with the automatically annotated galaxies, and in fact are in good agreement with the other experiments. Whether the agreement is the result of an astronomical reason or a certain unidentified human bias is still a matter that is difficult to fully determine due to the complex nature of the possible human biases of the crowdsourcing-based annotations.

\section{Analysis with Pan-STARRS data}
\label{panstarrs}

A sky survey that covers more of the Southern sky than SDSS is Pan-STARRS. Like SDSS, Pan-STARRS covers mostly the Northern sky, but its footprint covers more of the Southern hemisphere compared to SDSS. Using a dataset of Pan-STARRS DR1 galaxies used in a previous study \citep{shamir2017large} provided the magnitude difference of 3,587 galaxies in the $60^o \times 60^o$ around the Southern galactic pole. The galaxies can be accessed at \url{https://people.cs.ksu.edu/~lshamir/data/assym3/}. The results show that in the part of the sky around the Southern galactic pole galaxies that spin clockwise are brighter than galaxies that spin counterclockwise. That difference is in agreement with the results of DECam for the Southern galactic pole. The results are shown in Table~\ref{pan_starrs}.

\begin{table}
\caption{The g, r, and z exponential magnitudes of Pan-STARRS galaxies at around the Southern galactic pole.}   
\label{pan_starrs}
\scriptsize
\centering
\begin{tabular}{lccHcccc}
\hline
Band        & Mag           & Mag  & $ \Delta Mag$   & T-test P \\
               &  cw            &  ccw  & \\                                     & \\
\hline
G  &  16.9066$\pm$0.02      &  16.9972$\pm$0.02         & -0.0906  & 0.0014 \\
R  & 16.3463$\pm$0.02     &  16.4226$\pm$0.02       & -0.0763  &  0.007 \\
Z  & 15.8408$\pm$0.02      & 15.9087$\pm$0.02        & -0.0679   & 0.0164 \\
\hline
\end{tabular}
\end{table}


\section{HST data}
\label{hst}

As ground-based instruments, DECam, SDSS, and Pan-STARRS are subjected to the effect of the atmosphere. There is no known atmospheric effect that can effect galaxies differently based on their spin direction, and therefore the atmosphere is not expected to lead to such difference. To test empirically whether the difference is consistent also when imaging the galaxies without the effect of the atmosphere, we used galaxies imaged by the Hubble Space Telescope (HST). These results were shown initially in \citep{shamir2020asymmetry}.

The galaxies used in this experiment were imaged by the Cosmic Evolution Survey (COSMOS) of HST. As the largest HST field, COSMOS \citep{scoville2007cosmos,koekemoer2007cosmos,capak2007first} covers $\sim$2 square degrees, centered at ($\alpha=150.119^o$, $\delta=2.2058^o$). The initial list of objects included 114,630 COSMOS objects that are at least 5$\sigma$ brighter than their background. Each galaxy image was separated from the F814W image by using the {\it Montage} \citep{berriman2004montage} tool. The images were annotated by the Ganalyzer algorithm \citep{shamir2011ganalyzer}, and then inspected manually. That led to a dataset of 2,607 galaxies that spin clockwise, and 2,515 galaxies that spin counterclockwise. The full details of the annotation process are provided in \citep{shamir2020asymmetry}, and the annotated data can be accessed at \url{http://people.cs.ksu.edu/~lshamir/data/assym_COSMOS/}.

Table~\ref{comp1} shows the brightness in the g, r, and z filters of galaxies spinning in opposite directions. The magnitudes are the Subaru AB magnitudes \citep{capak2007first}. The comparison provides certain evidence that in the COSMOS field galaxies that spin counterclockwise are brighter than galaxies that spin clockwise. The COSMOS field is not aligned with neither the Northern nor the Southern galactic pole, but it is far closer to the Northern galactic pole. According to the other experiments described earlier in this paper, in the Northern galactic pole galaxies that spin counterclockwise are expected to be brighter. That observation is aligned with the results observed with the HST galaxies, reported in \citep{shamir2020asymmetry}. The difference in the z band is not necessarily statistically significant, but the findings are aligned with the results shown by the other telescopes, and definitely do not conflict with DECam, SDSS and Pan-STARRS.


\begin{table}[h]
\begin{center}
\caption{The brightness of HST galaxies spinning in opposite directions in the COSMOS field.}
\label{comp1}
\scriptsize
\begin{tabular}{lcccc}
\hline
Band & Mag cw & Mag ccw & $\Delta$Mag & P (t-test)  \\
\hline
G & 23.131$\pm$0.019  & 23.077$\pm$0.019 & 0.054 & 0.023 \\
R & 22.266$\pm$0.019 & 22.218$\pm$0.02 & 0.048   & 0.045  \\
Z & 21.358$\pm$0.017 & 21.323$\pm$0.018 & 0.035  & 0.087 \\
\hline
\end{tabular}
\end{center}
\end{table}

\section{Reasons that can lead to differences in brightness}
\label{error}

The differences in brightness between galaxies spinning in opposite directions cannot be determined by direct observation, but by analysis of a large number of galaxies. Large-scale analysis of a high number of galaxies to determine properties that are difficult to obtain with direct measurements is not a new practice, and is used in tasks such as weak gravitational lensing \citep{van2000detection,wittman2000detection,mandelbaum2006galaxy,hirata2007intrinsic,abbott2016cosmology}. The purpose of this section is to review the analysis and possible errors that can lead to the observation.

\subsection{Incorrectly annotated galaxies}
\label{annotation_error}

One of the key aspects of the analysis shown here is the annotation of the galaxies by their spin direction. Two different algorithms were used in this study, as well as manually annotated galaxies using crowdsourcing, all show similar results.  For the computer-based annotations, the experiments were repeated after mirroring the galaxy images, leading to inverse results. Also, the brightness difference is inverse in the Northern galactic pole compared to the Southern galactic pole. A bias in the algorithm is expected to be consistent in both ends of the Galactic pole, rather than flip.

Also, if an algorithm that annotates the galaxies by their spin directions annotates some of the galaxies with wrong spin direction, the brightness difference $\Delta M$ at a certain part of the sky is determined by Equation~\ref{delta_M}

\begin{equation}
\label{delta_M}
\Delta M = ((1-e)\bar{M_{cw}} + e\bar{E_{cw}}) - ((1-e)\bar{M_{ccw}} + e\bar{E_{ccw}}),
\end{equation}
where $\bar{M_{cw}}$ is the mean magnitude of galaxies spinning clockwise that are also annotated correctly by the classifier as galaxies spinning clockwise, $\bar{E_{cw}}$ is the mean magnitude of galaxies spinning counterclockwise but annotated incorrectly as spinning clockwise, and $e$ is the error rate of the annotation algorithm. The equation can be re-written as

\begin{equation}
\label{delta_M1}
\Delta M = \bar{M_{cw}} - \bar{M_{ccw}} + e(\bar{E_{cw}} - \bar{E_{ccw}} + \bar{M_{ccw}} - \bar{M_{cw}}).
\end{equation}

Let $\vartheta  = \bar{M_{cw}} - \bar{M_{ccw}}$ and $\varphi = \bar{E_{cw}} - \bar{E_{ccw}}$. $\Delta M$ can be now expressed as $\Delta M = \vartheta + e(\vartheta-\varphi)$. If the magnitude between clockwise and counterclockwise galaxies is different, it should be consistent for all galaxies, including galaxies that are annotated incorrectly. In that case, $\varphi=-\vartheta$, and $\Delta M=\vartheta+e(-2\vartheta)$. That shows that the $\Delta$M that includes galaxies annotated incorrectly is smaller than $\vartheta$. Therefore, a higher rate of error in the annotation $e$ will lead to a smaller $\Delta M$. Therefore, error in the annotation can lead to lower difference in magnitude, but since $e\geq0$, it cannot lead to a higher difference.

\subsection{Cosmic variance}
\label{cosmic_variance}

Galaxies as observed from Earth are not distributed in the sky in a fully uniform manner, leading to subtle fluctuations in galaxy density known as ``cosmic variance'' \citep{driver2010quantifying,moster2011cosmic}. These small fluctuations in galaxy population density can affect measurements at different parts of the sky and different directions of observation \citep{kamionkowski1997getting,amarena2018impact,keenan2020biases}. 

The probe used in this study is the brightness difference between galaxies imaged in the same part of the sky, by the same telescope, in the same exposure, and the same analysis methods. That is, anything that might affect the brightness of galaxies that spin with the Milky Way is also expected to affect galaxies spinning in the opposite direction. There is no attempt to compare magnitudes measured in two different parts of the sky, two different instruments, or even two different exposures. In all experiments the mean brightness of galaxies spinning in one direction is compared to the mean brightness of galaxies spinning in the opposite way such that all galaxies are in the exact same part of the sky. Any cosmic variance that might affect galaxies spinning with the Milky Way is expected to affect the mean magnitude of galaxies spinning in the opposite direction.

\subsection{Bias in the hardware or photometric pipelines of digital sky surveys}

Digital sky surveys are some of the more complex research instruments of our time. That complexity makes it very difficult to inspect every single part of these systems and ensure that no bias exists. At the same time, it is also difficult to propose a certain possible flaw that can lead to differences in the brightness of galaxies that spin in the same direction as the Milky Way and galaxies that spin in the opposite direction. Moreover, the difference in brightness flips between the Northern and Southern galactic poles, making it more difficult to propose an explanation based on a possible hardware or software flaw.

In this study, several different digital sky surveys were used, and the results are consistent across the different telescopes. While it is challenging to propose of a specific flaw in a digital sky survey that exhibits itself in such form in a single telescope, it is more difficult to think of such flaw in several unrelated sky surveys.

\subsection{Atmospheric effect}

Atmospheric effect might change the brightness of galaxies as observed from Earth, and can lead to differences in brightness observed in different parts of the sky. As also discussed in~\ref{cosmic_variance}, the comparison between the magnitudes are done in the same part of the sky, and all galaxies were taken from the exact same frames and exposures. Therefore, all atmospheric effects that can change the magnitude of galaxies that spin in one direction are expected to change the magnitude of galaxies that spin in the opposite direction.

To completely eliminate the atmospheric effect, an experiment was done with data from Hubble Space Telescope. That experiment is described in Section~\ref{hst}. The results of that experiment are consistent with the results from the ground-based telescopes, providing another indication that the difference in brightness is not the result of the effect of the Earth's atmosphere.

\subsection{Spiral galaxies with leading arms}

Although the vast majority of spiral galaxies have trailing arms, in some less common cases the arms of a spiral galaxy can be leading. For instance, a notable case of a galaxy with leading spirals arms is NGC 4622 \citep{freeman1991simulating,buta2003ringed,byrd2007ringed}. Assuming that the spin direction of a galaxy is driven by the perspective of the observer, the frequency of galaxies with leading arms among galaxies that spin in the same direction as the Milky Way is similar to the frequency of galaxies spinning in the opposite direction. In that case, galaxies with leading arms can be considered as galaxies that were annotated incorrectly, and subjected to the same analysis shown in Section~\ref{annotation_error}.


\section{Possible link to $H_o$ tension}
\label{H0}

The Hubble-Lemaitre constant ($H_o$) tension \citep{wu2017sample,mortsell2018does,bolejko2018emerging,davis2019can,pandey2020model,camarena2020local,di2021realm,riess2022comprehensive} is one of the most puzzling cosmological observations. In summary, the Hubble-Lemaitre constant $H_o$ determined by using Ia supernovae to measure the distance of galaxies is different from the constant measured with the cosmic microwave background radiation. Since both measure the same Universe, and assuming the Universe can only have one expansion rate, one of the explanations to the tension is that one of these probes has slight inaccuracies that lead to the different $H_o$ values. For instance, it has been suggested that Lorentz Relativistic Mass can affect the measurements using Ia supernovae \citep{haug2022does}. 

Hubble-Lemaitre constant can be determined by comparing the velocity of galaxies relative to their distance, such that the distance is determined by using Ia supernovae. Ia supernovae have an expected absolute magnitude, and therefore the apparent magnitude of an Ia supernova as observed from Earth can be used to determine the distance of the galaxy from Earth.

Supernovae are explosions of stars, and therefore a supernova rotates with the galaxy that the star was part of. The magnitude of the Ia supernova is measured from Earth, which also rotates with the Milky Way galaxy at around 220 km $\cdot$ sec $^{-1}$. If the rotation of the galaxies that host Ia supernovae compared to the rotation of the Milky Way can affect the apparent magnitude of the supernova, that can lead to a different distance metrics from Ia supernovae that depends on the rotation of the galaxy compared to the rotation of the Milky Way. That is, because not all supernovae used in the measurements are located around the galactic pole and spin in the same direction as the Milky Way, their apparent magnitude might seem slightly different to an Earth based observer, leading to a slight change to the measured Hubble-Lemaitre constant. Because the rotational velocity of a galaxy correlates with the galaxy type, the link between rotational velocity and $H_o$ can be related to previous reports on correlation between $H_o$ and the type of the Ia supernova host galaxy \citep{refId0}. Other studies suggested a link between Ia supernova and the properties of its host galaxies \citep{10.1093/mnras/stac3056,10.1093/mnras/stac2994}.

An experiment that would test that assumption can be made by computing the Hubble-Lemaitre constant by using only Ia supernovae hosted by galaxies that rotate in the same direction as the Milky Way. Such galaxies can be from both the Northern and Southern hemisphere, such that the galaxies are located around the galactic pole. That would require a certain number of Ia supernovae located around the galactic pole, and galaxies spinning in the same direction as the Milky Way. If the measurement of the Hubble-Lemaitre constant when using galaxies that rotate with the Milky Way provides a different result than when using all galaxies, that can provide an indication that the spin direction affects the determined Hubble-Lemaitre constant. If the result is also close to the Hubble-Lemaitre constant as determined by the CMB radiation, that can also explain the $H_o$ tension, and might also explain other observation made with Ia supernovae as distance candles. 

To perform a simple test, we used the code and data provided by \cite{refId0} for determining the $H_o$ constant. \cite{refId0} use a set of 140 Ia supernovae, calibrated with Pan-STARRS. The calibration is done using two sets. The primary one is a set of 24 {\it SBF} \citep{refId0} Ia supernovae. Additionally, a set of 19 Ia SH0ES \citep{riess2019large} supernovae is also used. The full description of the data and calibration is provided in \citep{refId0}. 

In addition to using the 140 supernovae as done in \citep{refId0}, we also separate a subset of the 140 supernovae into supernovae within 45$^o$ from the Southern galactic pole and their host galaxies spin clockwise, or within 45$^o$ from the Northern galactic pole and their host galaxies spin counterclockwise. That provided a list of 26 supernovae. Five of these supernovae had redshift of $z<0.02$, and were excluded from the analysis as was done in \citep{refId0}. The 21 supernovae are SN2005iq, SN2008Y, SN2004L, SN2006cq, SN2006or, SN2008bz, SN2005hc, SN2005eq, SN2008ar, SN2001ah, SN2006et, SN2002G, SN2005lu, SN1996bl, SN1999ef, SN2009na, SN2009D, SN2007O, SN1994Q, SN2006ac, SN2006cj.

When computing $H_o$ using the 140 supernovae and the SBF calibration, the $H_o$ constant is 69.070 km/s/Mpc. That was done by using the code in the file {\it Hierarchical\_noHM\_SBF} \footnote{\url{https://github.com/nanditakhetan/SBF_SNeIa_H0}}. The 3\% error range of (64.747, 73.622). When using the subset of 21 supernovae, the computed $H_o$ increases to 69.551 km/s/Mpc, and 3\% error range of (63.880, 75.787). That change makes the $H_o$ closer to the $H_o$ measured with the CMB. Due to the large error margins the change is not statistically significant, but it is aligned with the contention that normalization of the spin directions can reduce the Hubble tension. 

The observed $H_o$ when using the 21 supernovae with normalized spin and SBF calibration is not the same as the $H_o$ observed with CMB, and therefore does not provide an immediate explanation to the $H_o$ tension. That can be explained in several ways. The first is that the reason proposed here is not the real cause of the $H_o$ tension, and there are other factors that affect the measurements. Another reason is that the Ia supernovae used here are within certain proximity to the galactic pole, but not exactly on it. Also, the declination of these galaxies is not zero, which can also lead to a certain Doppler shift effect. A more suitable future experiment would require a larger number of Ia supernovae that are exactly on the galactic pole, and are full face-on galaxies. Still, the $H_o$ observed with the CMB radiation is within the error range of the $H_o$ determined here.

\section{Conclusions}
\label{conclusions}

The results show that in the fields around the galactic pole galaxies that spin clockwise have a different brightness than galaxies that spin counterclockwise. Such differences are expected due to relativistic beaming of galaxies that spin in the opposite direction to that of the Milky Way. The inverse difference in the opposite ends of the galactic pole shows that the observations in the both ends are in agreement with the spin direction of the Milky Way. Clearly, a galaxy that seem to spin clockwise in the Northern galactic pole spins in the same direction as galaxies that seem to spin counterclockwise in the Southern galactic pole, and vice versa. The control fields that are 90$^o$ from the galactic pole show no significant difference between galaxies spinning in opposite directions, which provides a certain indication of a correlation between the galactic pole and the magnitude difference.  The contention that relativistic beaming can lead to a certain statistical asymmetry when observing a population of galaxies is not unexpected \citep{alam2017relativistic}.


Future and more powerful telescopes such as  the Vera Rubin Observatory will allow better profiling of such difference. In case the maximum magnitude difference is greater than the maximum expected difference under optimal conditions, that observation can be related to modified Newtonian dynamics, or to relativistic beaming of gravity as an explanation to the galaxy rotation curve anomaly \cite{blake2021relativistic}. It can also be related to the observation of disagreement between the observed and expected velocity of halo spin \cite{libeskind2013velocity}.






The observed difference between the magnitude of galaxies that spin in opposite directions had also been observed in previous studies \cite{shamir2016asymmetry,shamir2017large,shamir2017photometric,shamir2020asymmetry}, showing differences between galaxies that spin in opposite directions in the same field. These studies were based on far smaller datasets collected by sky surveys that did not cover the fields around both galactic poles. The data collected by DECam and available through the DESI Legacy Survey allows direct comparison of the differences in brightness between the two ends of the galactic pole. These differences can be related to the effect of relativistic beaming, but further research will be required with more powerful instruments to quantify and profile the magnitude difference.

\subsection{Large-scale structure explanation}

Another possible explanation to the observation can be an anomaly in the large-scale structure of the Universe. That is, the axis observed around the galactic pole is not the result of differences in the galaxy brightness as observed from Earth, but the result of the alignment in the spin directions of galaxies in the Universe. That explanation requires modification to the standard cosmological models, which rely on the assumption that the Universe is homogeneous and isotropic, an assumption defined as the {\it Cosmological Principle}.

While the Cosmological Principle is a common working assumption for most cosmological models, it has not been fully proven. In fact, multiple probes have shown evidence of violation of the cosmological principle \citep{aluri2022observable}. Perhaps the most notable probe is the cosmic microwave background radiation, also exhibiting a cosmological-scale axis \citep{abramo2006anomalies,mariano2013cmb,land2005examination,ade2014planck,santos2015influence,gruppuso2018evens,yeung2022directional}. It has been suggested that the axis exhibited by the cosmic microwave background radiation agrees with other axes formed by probes such as dark energy and dark flow \citep{mariano2013cmb}. Other anomalies related to the cosmic microwave background radiation are the quandrupole-octopole alignment \citep{schwarz2004low,ralston2004virgo,copi2007uncorrelated,copi2010large,copi2015large}, the asymmetry between hemispheres \citep{eriksen2004asymmetries,land2005examination,akrami2014power}, and the point-parity asymmetry \citep{kim2010anomalous,kim2010anomalous2}. Another related observation is the CMB cold spot \citep{cruz655non,masina2009cold,vielva2010comprehensive,mackenzie2017evidence,farhang2021cmb}. It has also been suggested that the isotropy observed with the CMB radiation is not statistically significant \citep{bennett2011seven}.

In addition to the CMB, probes that show anisotropy include radio sources \citep{ghosh2016probing,tiwari2015dipole,tiwari2016revisiting,Singal2019,marcha2021large}, LX-T scaling \citep{migkas2020probing}, short gamma ray bursts \citep{meszaros2019oppositeness}, acceleration rates \citep{perivolaropoulos2014large,migkas2021cosmological,krishnan2022hints}, Ia supernova \citep{javanmardi2015probing,lin2016significance}, distribution of galaxy morphology types \citep{javanmardi2017anisotropy}, dark energy \citep{adhav2011kantowski,adhav2011lrs,perivolaropoulos2014large,colin2019evidence}, fine structure constant \citep{webb2011indications}, galaxy motion \citep{skeivalas2021predictive}, $H_o$ \citep{luongo2022larger}, polarization of quasars \citep{hutsemekers1998evidence,hutsemekers2005mapping,secrest2021test,zhao2021tomographic,semenaite2021cosmological}, and cosmic rays \citep{sommers2001cosmic,deligny2013searches,aab2017observation,aab2019probing}. It has also been shown that the large-scale distribution of galaxies in the Universe is not random \citep{jones2005scaling}, and could have a preferred direction \citep{longo2011detection,shamir2012handedness,shamir2019cosmological,shamir2020patterns,shamir2020pasa,shamir2021large,shamir2022asymmetry,shamir2022analysis,philcox2022probing,hou2022measurement}. These probes might not agree with the standard models \citep{pecker1997some,perivolaropoulos2014large,bull2016beyond,velten2020hubble,krishnan2022hints,krishnan2022does,luongo2022larger,colgain2022probing,abdalla2022cosmology}

The contention of a cosmological-scale axis agrees with cosmological theories that shift from the standard models. For instance, black hole cosmology \citep{pathria1972universe,stuckey1994observable,easson2001universe,seshavatharam2010physics,poplawski2010radial,christillin2014machian,dymnikova2019universes,chakrabarty2020toy,poplawski2021nonsingular,seshavatharam2022concepts,gaztanaga2022black,gaztanaga2022black2} is a cosmological theory aligned with the contention of the existence of a large-scale axis. Black holes are born from the collapse of a star, and since stars spin black holes also spin \citep{gammie2004black,takahashi2004shapes,volonteri2005distribution,mcclintock2006spin,mudambi2020estimation,reynolds2021observational}. Supermassive black holes are also expected to spin \citep{montero2012relativistic}, and observations of supermassive black holes showed that supermassive black holes spin \citep{reynolds2019observing}. An early example of such observation is the spin of the supermassive black hole of NGC 1365 \citep{reynolds2013black}. Since black holes spin, if the Universe is the interior of a back hole it is expected to spin around a major axis inherited from the black hole. That observation is aligned with the agreement between the Hubble radius of the Universe and the Schwarzschild radius of a black hole such that the mass of the black hole is the mass of the Universe \citep{christillin2014machian}. Black hole cosmology is a theory under the category of multiverse \citep{carr2008universe,hall2008evidence,antonov2015hidden,garriga2016black}, which is one of the first cosmological paradigms \citep{trimble2009multiverses,kragh2009contemporary}. 

In addition to black hole cosmology, several other cosmological theories that assume the existence of a cosmological scale axis have been proposed. These theories include ellipsoidal universe \citep{campanelli2006ellipsoidal,campanelli2007cosmic,campanelli2011cosmic,gruppuso2007complete,cea2014ellipsoidal}, flat space cosmology \citep{tatum2015basics,tatum2015flat,tatum2018clues,azarnia2021islands}, geometric inflation \citep{arciniega2020geometric,edelstein2020aspects,arciniega2020towards,jaime2021viability}, supersymmetric flows \citep{rajpoot2017supersymmetric}, and rotating universe \citep{godel1949example,ozsvath1962finite,ozsvath2001approaches,su2009universe,sivaram2012primordial,chechin2016rotation,chechin2017does,seshavatharam2020integrated,camp2021}. Other theories are double inflation \citep{feng2003double}, f($\mathscr{R}$, $\mathscr{L}_m)$ gravity \citep{KAVYA2022101126}, contraction prior to inflation \citep{piao2004suppressing}, primordial anisotropic vacuum pressure \citep{rodrigues2008anisotropic}, moving dark energy \citep{jimenez2007cosmology}, multiple vacua \citep{piao2005possible}, and spinor-driven inflation \citep{bohmer2008cmb}. While these theories shift from the standard cosmological models, they provide an explanation to an observed axis of a cosmological-scale. Since the standard model has not been fully proven, alternative theories to the standard cosmology should also be considered.

\section{Discussion}
\label{discussion}

Despite decades of intensive research efforts, the physical nature of galaxy rotation is still an unsolved question. The common possible explanations are that the distribution and quantity of the mass of galaxies does not fit its physical properties (dark matter), or that the laws of physics are different when applied to galaxies. This paper proposes the contention that the rotational velocity of the galaxy does not fit its physical properties, and corresponds to a much higher rotational velocity. The explanation is driven by the observation that galaxies that spin in the same direction as the Milky Way have different brightness compared to galaxies that spin in the opposite direction. 

While such difference in brightness can be explained by Doppler shift, the difference is expected to be small. The observed difference is far greater than the expected difference, and it is observable with Earth-based telescopes. A possible explanation is that the physics of galaxy rotation corresponds to a far higher rotational velocity than the observed velocity. While that explanation is physically provocative, the physics of galaxy rotation is still unexplained, and does not follow known or proven physical theories. For instance, the absence of Keplerian velocity decrease in the galaxy rotation curve was ignored for several decades, also because it did not agree with the physical theories \citep{rubin2000one}. 

The tension between galaxy rotational velocity and its physical properties is one of many other possible explanations to the observation of different brightness for galaxies spinning in opposite directions. Another possible explanation discussed here is a possible anomaly in the large-scale structure. In that case, the difference in brightness is not driven by the perspective of an Earth-based observer, but reflects the real structure of the Universe. Other explanations not considered in this paper can also be possible. 

The link between the rotational velocity of a galaxy and the apparent brightness of its objects can also be related to measurements made with Ia supernovae. Measurements with Ia supernovae have provided unexpected results, and have been shown to be in disagreement with other measurements. For instance, the $H_o$ tension suggests that the CMB or Ia supernovae do not match, and since both are applied to the same Universe it can be assumed that one of these measurements is inaccurate. If the brightness of an Ia supernovae depends on the rotational velocity of the galaxy that hosts it, and the galaxy may not necessarily spin in the same direction as the Milky Way, that can lead to slight changes in the apparent magnitude of the supernovae. Because the distance of an Ia supernovae is determined from its brightness, unexpected changes in the apparent magnitude can lead to a slightly different distance, and consequently to a slightly different $H_o$. The initial observation of the accelerated expansion of the Universe was also made with Ia supernovae, and such changes in brightness can also have a certain impact on these results.


\section*{Acknowledgments}

This study was supported in part by NSF grants AST-1903823 and IIS-1546079. 

\bibliography{galaxy_velocity}

\end{document}